\documentstyle[12pt]{article}
\begin{document}
\begin{flushright}
FERMILAB-Conf-94/249\\
hep-ex/9408005
\end{flushright}
\vspace{0.1cm}
\begin{center}
{\large\bf TEVATRON ENERGY AND LUMINOSITY UPGRADES BEYOND THE MAIN INJECTOR}\\
\vspace{0.5cm}
D. AMIDEI{\em$^c$}, A. BADEN{\em$^b$}, G. W. FOSTER{\em$^a$},
G.P. JACKSON{\em$^a$},
T. KAMON{\em$^d$}, J.L.~LOPEZ{\em$^d$}, P. MCINTYRE{\em$^d$},
J. STRAIT{\em$^a$}, and J. WHITE{\em$^d$}\\
\vspace{0.2cm}
{\em $^a$Fermi National Accelerator Laboratory} \\
{\em $^b$Department of Physics, University of Maryland}\\
{\em $^c$Department of Physics, University of Michigan}\\
{\em $^d$Department of Physics, Texas A\&M University }
\end{center}
\begin{abstract}
The Fermilab Tevatron will be the world's highest
energy  hadron collider  until the  LHC is commissioned, it has the world's
highest energy fixed target beams, and Fermilab will be the leading high energy
physics laboratory in the US for the foreseeable future. Following the demise
of the SSC, a number of  possible upgrades to the  Tevatron complex, beyond
construction of the Main Injector, are being discussed.  Using existing
technology, it appears possible to increase the  luminosity of the $\bar{p}p$
Collider to at least $10^{33}$cm$^{-2}$sec$^{-1}$ (Tevatron-Star) and to
increase the beam energy to 2 TeV (DiTevatron).  Fixed  target beam of energy
about 1.5 TeV could also be delivered. Leaving the existing Tevatron in the
tunnel and constructing bypasses around the collider halls would allow
simultaneous 800 GeV fixed target and $\sqrt{s}$ = 4 TeV collider operation.
These upgrades would give Fermilab an exciting physics program which would be
complementary to the LHC, and they would lay the  groundwork for the
construction of a possible post-LHC ultra-high energy hadron collider.
\end{abstract}

\footnotetext{\bf Presented at the Eighth Meeting of the Division of Particles
and Fields Albuquerque, New Mexico, August 2-6, 1994.}
\vspace{-0.3cm}
\section{Introduction}
Ideas for upgrading the energy and luminosity of the Tevatron beyond that
planned with the Main Injector are not new~\cite{Upgrades}.  However, the
cancellation of the SSC has given new impetus to developing and implementing
these ideas. Fermilab is the US flagship HEP lab, and if there is to be a
future for  hadron physics in the US, it will be centered at Fermilab.
Continued  development and enhancement of its facilities and physics research
program are necessary to maintain its long term vitality.  For the last decade
hadron accelerator physics has been concentrated on a single large construction
project, the SSC.  As a result, a number of promising R\&D
projects have been left at the wayside.  With the SSC gone, it is now
appropriate to resume efforts on a mixture of long-term research pointing
toward far-future facilities and short-term development of the capabilities of
the existing facilities.

There is important physics research which an upgraded Tevatron can
address. A $\sqrt{s}$ = 4 TeV, $\cal L$ = $10^{33}$/cm$^{2}$/s $\overline{p}p$
collider is a ``Top Factory,'' producing up to 1000 reconstructed
$t\overline{t}$ pairs per week.   The sensitivity to detect an intermediate
mass Higgs (80$\sim$120 GeV) via WH associated production with $H \rightarrow b
\overline{b}$ is comparable to that of LHC~\cite{Willenbrock}.  Most of the
parameter space of a set of well motivated constrained SUSY models can be
covered: gluino mass up to 750 (450) GeV/c$^{2}$ at $\sqrt{s}$ = 4 (2) TeV and
chargino mass up to 150--200 GeV/c$^{2}$ for either energy~\cite{Kamon}.
An energy upgrade is effectively a luminosity upgrade for fixed target
physics. Doubling the primary beam energy increases the flux of secondary
beams by up to an order of magnitude. Some upgrade scenarios involve leaving
the existing Tevatron in the tunnel, which would allow 800 GeV
fixed target running simultaneously with 2+2 TeV collider operation. Charm,
hyperon, and neutrino physics would be enhanced.  For example, experiments
with more than $10^{8}$ reconstructed charmed mesons are possible.

Physics done by the LHC and a $\cal L$ = 10$^{33}$/cm$^{2}$/s, $\sqrt{s}$ = 4
TeV $\bar {p}p$ collider are complementary. In the few hundred GeV mass range,
where we
expect there to be new phenomena, a $\sqrt{s}$ = 14 TeV pp collider is
dominantly a gg collider and  a $\sqrt{s}$ = 2-4 TeV $\overline{p}p$ collider
is dominantly a $q\overline{q}$ collider~\cite{Hill}.  For processes such as WH
associated production and $t\bar{b}$ production, which proceed dominantly from
a $q\bar{q}$ initial state and for which the dominant background comes from gg
interactions, the signal to background ratio will be better in a $\bar{p}p$
than a $pp$ collider.   If LHC concentrates on Higgs research to the highest
possible mass (or in the intermediate mass range in the $H \rightarrow \gamma
\gamma$ mode), it must run at the highest possible luminosity and therefore
with at least  15~interactions/crossing.  With the DiTevatron operating such
that there are only a few interactions per crossing, physics that relies on
b-tagging and good missing $E_T$ resolution can be done with greater
efficiency.    Fixed target beams up to $\sim$1.5 TeV will allow physics to be
done that  cannot be done at LHC (or elsewhere);  e.g. ultra-high statistics
charm physics, perhaps leading to the observation of $D^{0}\overline{D}\ ^{0}$
mixing and maybe even CP violation.

The LHC is unlikely to answer all the questions we know to ask now, nor does it
represent the limit of hadron collider technology.  Therefore, we will want to
build an ultra-high energy ($\geq$ 30+30 TeV) hadron collider after LHC. One
lesson of the SSC is that we should build incrementally on existing facilities,
rather than starting a new lab when we need a new accelerator, and  a 2 TeV
rapid cycling accelerator would be an appropriate injector for the post-LHC
machine.  The accelerator R\&D required for the energy and luminosity upgrades
of the Tevatron, for example the development of antiproton source technology or
the  generation and acceleration of high-brightness proton beams,  could help
us learn how to build future machines more economically.

Two potential upgrades are being considered, which may be implemented
separately or together. The first, Tevatron-Star (TeV*), is an increase in the
collider luminosity to $\cal L$ = 10$^{33}$/cm$^{2}$/s (10~fb$^{-1}$/yr),  an
order of magnitude  above that with the Main Injector.  The principal project
is to upgrade the $\bar{p}$ source to increase  $\overline{p}$ production by a
factor of 6 and to recycle  the remaining $\overline{p}$'s at the end of a
store.  Improvements to the injector chain would also be needed to produce a
brighter proton beam.   The second, the DiTevatron, uses  SSC magnet technology
to double the Tevatron energy to 2~TeV per beam.  Adiabatic damping of the beam
size would give a natural $\times$2 increase in luminosity over the Tevatron to
$\cal L$ = 2$\times10^{32}$/cm$^2$/s  (2~fb$^{-1}$/yr) without the TeV* upgrade
and $\cal L$ = 2$\times10^{33}$/cm$^2$/s  (20~fb$^{-1}$/yr) with it.

\section{Tevatron-Star}

Currently, one Booster batch (84 bunches) in the Main Ring (and later the Main
Injector) is targeted for $\overline{p}$ production.  Antiprotons are collected
with a 4\% momentum bite and cooled and stored in 2 rings -- the Debuncher and
Accumulator -- which  operate at the same energy (8 GeV) and have a similar
circumference as the Booster.  At the end of a store the remaining
$\overline{p}$'s in the Tevatron, about 30\% of the initial number, are dumped.
With the Main Ring, the stacking rate is 4$\times$10$^{10}$/hour and with the
Main Injector it is expected to increase to  15$\times$10$^{10}$/hour.  If the
$\bar{p}$'s remaining at the end of the store can be re-used and a factor of 6
higher stacking rate is obtained, then an order of magnitude luminosity
increase
will be realized.

The ``simplest'' method to increase the $\bar{p}$ production rate  is to target
6 Booster batches (the capacity of the Main Injector allowing for kicker gaps)
rather than one.  This requires the construction of a new ring, the Compressor,
which has the same circumference as the Main Injector and which would be placed
in the same tunnel.  Bunch rotation would be performed in the new ring, and
then the 6 Booster batches worth of $\bar{p}$'s would be compressed azimuthally
and transferred to the existing  Antiproton Source for cooling and
accumulation.  A second new ring, the Recycler, is required to store the stack
of at least  1.3$\times$10$^{13}$, which substantially exceeds the capacity of
the existing  Accumulator.   The Recycler would have only a core cooling
system and would operate above the Main Injector transition energy.  It would
also be used to accept and recool the $\bar{p}$'s remaining in the Tevatron at
the end of the store.  The Recycler  could be located in either the Main
Injector or Antiproton Source tunnel.

Table 1 compares several possible luminosity and energy upgrade scenarios. In
all cases the following parameters are assumed: $\beta$*=25~cm, recovered
$\bar{p}$ fraction = 30\%, $\bar{p}$ transfer efficiency = 80\%, coalescing
efficiency = 75\%, and stacking time between stores = 13 hours.  The first
column gives the expected performance with the Main Injector.  For 1(2) TeV
beam and 36 bunches, $\cal L$ = 1.2(2.4)$\times$10$^{32}$/cm$^2$/s and there
will be about 3(6) interactions per crossing.  If the luminosity is increased
ten-fold, the number of interactions per crossing would become unacceptably
large.  The second column (TeV*~A) is the case in which the number of bunches
is increased to 108 and the  spacing is reduced to 132 ns.  The collider
detector electronics are currently being upgraded to accommodate this bunch
spacing.  It is assumed that modest improvements in the injector chain, either
improved operation of the existing accelerators or accelerator upgrades, will
result in a 40\% decrease in the proton beam emittance and a 60\% increase in
the number of protons per RF bucket, from 40$\times$10$^9$ to 64$\times$10$^9$.
With these parameters the desired 1(2)$\times$10$^{33}$ luminosity is realized
at 1(2)~TeV per beam, and there will be 9(17)  interactions per crossing.

The case TeV* B uses uncoalesced beams to decrease the number
of interactions per crossing.  Due to the smaller momentum spread of the
uncoalesced beam, the bunch length decreases, giving an increase in the ``hour
glass'' form factor.  However, a non-zero crossing angle, such that the beams
are separated by 5$\sigma$ at the next crossing point
reduces the overall form factor.  With lower density bunches, the luminosity
is now about 1/3 of the desired value.  The last column (TeV* C) assumes that
the emittances can be reduced to 1 $\pi$ mm-mrad by appropriate accelerator
improvements.  This is the most desirable case: $\cal L$
=1-2$\times$10$^{33}$/cm$^2$/s, and 1-3 interactions per crossing.

\begin{table}[tb]
\caption{Tevatron Upgrade Scenarios}
\begin{center}
\begin{tabular}{|l|c|c|c|c|}
\hline
\hline
& \multicolumn{1}{|c|}{Main Injector} &
\multicolumn{1}{|c|}{TeV* A} & \multicolumn{1}{|c|}{TeV* B} &
\multicolumn{1}{|c|}{TeV* C}\\
\hline
$\bar{p}$ Production Rate ($10^{10}$/hour) & 15 & 90 & 90 & 90\\
$E_{beam}$ (TeV) & 1(2) & 1(2) & 1(2) & 1(2)\\
Number Coalesced & 13 & 5 & 1 & 1\\
Number of Bunches & 36 & 108 & 750 & 750\\
Bunch Spacing (ns) & 395 & 132 & 19 & 19\\
$N_{p}$/bunch ($10^{9}$) & 390 & 240 & 64 & 64\\
$N_{\bar{p}}$/bunch ($10^{9}$) & 33 & 93 & 18 & 18\\
$\epsilon_{p}$ (rms, $\pi$mm mrad) & 5 & 3 & 3 & 1\\
$\epsilon_{\bar{p}}$ (rms, $\pi$mm mrad) & 2.5 & 2.5 & 2.5 & 1\\
rms Bunch Length (cm) & 45 & 30 & 10 & 8\\
Crossing Half Angle (mrad) & 0 & 0 & 0.27 & 0.15\\
Luminosity Form Factor & 0.6 & 0.7 & 0.64 & 0.77\\
\hline
Peak ${\cal L}$ $10^{32}$/cm$^{2}$/s & 1.2(2.9) & 10(20) & 3(6) & 11(21)\\
Integrated ${\cal L}$ ($pb^{-1}$/week) & 24(48) & 200(400) & 65(130) &
220(430)\\
Interactions/crossing (45 mb) & 3.1(6.2) & 9(17) & 0.4(0.8) & 1.3(2.7)\\
\hline
p Tune Shift & 0.003 & 0.009 & 0.002 & 0.004\\
$\bar{p}$ Tune Shift & 0.019 & 0.020 & 0.005 & 0.016\\
\hline
\hline
\end{tabular}
\end{center}
\end{table}

There are many technical issues that must be addressed to reach the parameters
of the TeV* C case.  To use 6 Booster batches without destroying the target
requires a more advanced beam sweeping system (hard), defocussing the beam
(reduced $\bar{p}$ production efficiency) or use of multiple target
stations (expensive).  Bunch rotation in the Compressor is done most easily
near transition, or at $\sim$15 GeV.  This  requires deceleration either in the
Compressor (complicates the Compressor design) or the Main Injector (increased
cycle time).  Pre-cooling in the large circumference Compressor ring requires
impractically large power, and to cool the larger $\bar{p}$ flux, the
Accumulator
cooling system bandwidth would have to be increased by  a factor of eight.  To
generate the very low emittance beams probably requires replacing the
present source with an RFQ, and preserving the low emittance through
all of the accelerators will be challenging.

The very small beams required for high luminosity with uncoalesced beam may
have unacceptably short  emittance growth times due to intra-beam scattering.
Experiments are planned at Fermilab to measure this effect.  The large number
of  long-range (``parasitic'') beam-beam collisions (1500) may give rise to
unacceptable tune spread and tune shift.  The use of small, low emittance beams
tends to mitigate this problem since the tune shift and tune spread are
proportional to (D/$\sigma)^{-2}$ and (D/$\sigma)^{-3.8}$
respectively~\cite{SiemannJackson}, where $\sigma$ is the rms beam size and D
is
the beam separation.  (The achievable values of D/$\sigma$ will be discussed
further below.)  This effect can be studied in the Tevatron by injecting a
bunch of $\bar{p}$'s along with a fixed-target proton beam.

\section{DiTevatron}

Upgrading the beam energy to 2~TeV requires the construction of a new
accelerator with magnets twice the strength of Tevatron magnets.   Assuming the
same lattice as the Tevatron, dipoles of 8.8~T and arc quadrupoles of 150~T/m
are required.  Preliminary designs exist in which the IR quadrupole gradient
would only be increased to 210~T/m.  All of these magnet strengths are
achievable using technology developed for the SSC.  The energy upgrade could be
implemented alone or together with the luminosity upgrade discussed above.

Two main variants of this proposal under discussion: 1) Remove the  Tevatron
and inject into the DiTevatron directly  from the Main Injector.  High energy
fixed target beams would be available only when the collider was not running
and 120~GeV fixed target beam would be available during collider runs only if
the luminosity upgrade does not require the full Main Injector beam.   2) Leave
the Tevatron in the tunnel (or re-assemble it above the DiTevatron) and use it
as a high energy injector.  If bypasses were built around the collider
experiments, 800 GeV fixed target beams could be delivered during a collider
run. This would also allow the possibility, not considered further here, of 1+2
TeV pp collisions.  However, the expense of operating two superconducting
accelerators would be substantial.

The Tevatron collider runs about 5\% below the point where the weakest magnet
quenches.  Therefore, DiTevatron magnets must reach 9.2 T to
allow reliable 2 TeV collider operation.  The SSC dipoles provide a ``proof of
principle'' that the required field strength can be  achieved.  Twenty
full-scale SSC dipoles built at Fermilab and BNL routinely reached 7.2~T at
4.35~K and 8 T at 3.5~K~\cite{ASST}.  One magnet was tested at Fermilab at
1.8~K; it reached 9.4~T on the first quench and 9.6~T after 5
quenches~\cite{PostASST}.

Tevatron dipoles, which have a coil inner diameter of 76~mm, have a ``good
field region'' (over which  $\delta B$/$B_{0}$ $\leq$ 10$^{-4}$) of about 50~mm
diameter.  SSC dipoles have a 50~mm aperture, and the early models had only a
15~mm good field
region~\cite{Strait}.  With minor modifications to the design this could be
increased to about 35~mm, which may be large enough to accommodate the smaller
beams associated with the luminosity upgrade.

The largest aperture for collider operation is required at injection.  In
current 6 bunch operation the beams are separated from each other by 5$\sigma$
$\cong$ 15 mm.  If both beams are to be $\geq 5\sigma$ from the edge of the
good field region, a good field aperture $\geq$ 45~mm is required,  just less
than that in the Tevatron magnets and larger than can be achieved with SSC
dipoles.
Table 2 compares the beam size at  150~GeV in the Tevatron with those  for the
upgrade configurations given in Table 1.  Lattices have been worked out for an
energy upgrade that have smaller maximum dispersion than the
Tevatron~\cite{GarrenSyphers}.  This assumed in the last 4 columns.  The value
of D/$\sigma$ in the last row results from putting  the beams 5$\sigma$ from
the edge of the 50 (35)~mm good field region  in the Tevatron (DiTevatron).  In
the case TeV*~C, the possible separation is $>$ 30~$\sigma$, which may be
sufficient to keep the parasitic beam-beam effects under control.

\begin{table}[htb]
\caption{DiTevatron Beam Sizes}
\begin{center}
\begin{tabular}{|l|c|c|c|c|c|}
\hline
\hline
& \multicolumn{1}{|c|}{Tevatron} & \multicolumn{4}{|c|}{DiTevatron}\\
\hline
& \multicolumn{1}{|c|}{Main Injector} & \multicolumn{1}{|c|}{Main Injector} &
\multicolumn{1}{|c|}{TeV* A} & \multicolumn{1}{|c|}{TeV* B} &
\multicolumn{1}{|c|}{TeV* C}\\
\hline
$\epsilon_{p}$ (rms, $\pi$mm mrad) & 5 & 5 & 3 & 3 & 1\\
$\epsilon_{\bar{p}}$ (rms, $\pi$mm mrad) & 2.5 & 2.5 & 2.5 & 2.5 & 1\\
$\sigma_p/p (10^{-4})$ & 5 & 5 & 2.4 & 0.9 & 0.9\\
Dispersion (m) & 5 & 3.6 & 3.6 & 3.6 & 3.6\\
\hline
$\sigma_{xp}$ (mm) & 3.1 & 2.5 & 1.6 & 1.4 & 0.9\\
$\sigma_{x\bar{p}}$ (mm) & 2.8 & 2.2 & 1.5 & 1.3 & 0.9\\
D/$\sigma_{xp}$ & 6 & 4 & 12 & 15 & 31\\
\hline
\hline
\end{tabular}
\end{center}
\end{table}

The existing refrigeration system has the capacity at 4.5 K to remove heat from
the Tevatron due to the sum of the cryostat and power lead heat leaks
plus AC losses during fixed target running.
DiTevatron magnets would have a much more efficient cryostat, but would operate
at 1.8 K.  The greater cost of removing heat from 1.8 K than 4.5 K ($\times$6)
would roughly cancel the lower  heat leak of the DiTevatron,  so the capacity
of the existing refrigeration system would be adequate for the DiTevatron as a
collider~\cite{Theilacker}.   However, AC losses for fixed target operation
would be substantially larger than for the Tevatron due to the larger energy
swing, the larger volume of superconductor, and eddy current losses in the cold
iron yoke.  The lower refrigeration efficiency at 1.8 K would make 2~TeV fixed
target operation an order of magnitude more expensive than current 800 GeV
running and would require a correspondingly larger refrigerator.  Thus
2~TeV fixed target operation seems impractical.

Allowing a  15\% quench margin, as with the Tevatron in fixed target
mode, the DiTevatron could operate to 1.4 (1.6)~TeV at 4.4 (3.5)~K.
The AC losses for 1.5 TeV are only about 70\% as large as for 2~TeV, and the
refrigeration efficiency is larger at higher temperature.   Also, the limited
length of the straight sections in the existing tunnel (50~m) makes extraction
at 2 TeV much more difficult than at 1.5 TeV~\cite{Murphy}.  Thus 1.4--1.6~TeV
appears to be the practical limit for fixed target beam from the DiTevatron.

For secondary fixed target beams, an energy increase of the primary  beam is
effectively a luminosity increase.  An alternate way to deliver more integrated
luminosity would be to leave the Tevatron in the tunnel, build bypasses around
the collider experiments, and operate the 800~GeV fixed target program a larger
fraction of the time than is now possible.  It is  desirable that the bypasses
be in the same plane as the Tevatron.  Using SSC-strength magnets, operated at
4.4 K with a 15\% margin, a bypass that separates the Tevatron and DiTevatron
beams  by 10 m at the collision points would occupy about 7\% of the
circumference of the Tevatron and require about thirty 12 m long dipoles per
interaction region.

\section{Conclusions}

The energy and luminosity upgrades to the Fermilab accelerator complex
discussed in this paper can be achieved with existing technology, although the
state-of-the-art will be pushed in some cases.  There are strong reasons, both
for physics and for the health of the US HEP program, to invest  in Fermilab
beyond the Main Injector:  Important physics, which is complementary to the
LHC, can be done with a $\sqrt{s}$ = 4 TeV, $\cal L$
=2$\times$10$^{33}$/cm$^{2}$/s
$\overline{p}p$  collider and with enhanced fixed target beams (1.5 TeV between
collider runs or 800 GeV ``full time'').  An upgraded accelerator complex at
Fermilab would be a first-rate injector into a possible post LHC ultra-high
energy hadron collider.  And the investment in  accelerator R\&D could lead
to new developments that could make such an ultra-high energy collider less
expensive to build.  Further study will continue to develop these ideas and
plans.

\section*{Acknowledgements}

We would like to thank the following people who have contributed ideas and
calculations to the Tevatron upgrade proposals discussed here.  G.~Annala,
S.~Caspi, J.~Conrad D.~Finley, J.~Fuerst, R.~Gupta, C.~Hill, M.~Johnson,
D.~Kaplan, J.~MacLachlan, J. Marriner, E.~McCrory, C.T.~Murphy, J.~Theilacker,
and S.~Willenbrock.  We apologize to others who have also contributed but whom
we forgot to list.  This work was supported by the US Department of Energy.

\bibliographystyle{unsrt}

\begin{thebibliography}{99.}
\vspace{-.125in}

\bibitem{Upgrades} See, for example, M.J. Syphers and D.A. Edwards,  ``A $\geq$
1.5~TeV Superconducting Synchrotron Design for the Fermilab Tunnel,'' {\em
Proceedings of the 1989 IEEE Particle Accelerator Conference} (1989) 1821;  S.
Holmes, G. Dugan, and S. Peggs,  ``Ultimate Luminosity in the Tevatron
Collider,'' {\em Proceedings of the 1990 Summer Study on High Energy Physics}
(1990) 674.

\bibitem{Willenbrock}A. Stange, W. Marciano, and S. Willenbrock,  ``Associated
Production of Higgs and Weak Bosons, with $H\rightarrow b \bar b$, at Hadron
Colliders,''  ILL-(TH)-94-8 (1994); S. Mrenna and G.L. Kane,  ``Possible
Detection of a Higgs Boson at Higher Luminosity Hadron Colliders,'' CIT 68-1938
and UM-TH-94-24 (1994), submitted to {\em Phys. Rev. Lett.}.

\bibitem{Kamon}T. Kamon, J.L. Lopez, P. McIntyre and J.T. White,
``Supersymmetry at the DiTevatron,''  CTP-TAMU-19/94, to be published in  {\em
Phys. Rev.} {\bf D} (1994)

\bibitem{Hill}C. Hill, ``The Future of Top Physics,'' Presented at the Fermilab
Users Annual Meeting, June 9-10, 1994.

\bibitem{SiemannJackson}R.H. Siemann and G.P. Jackson, ``A High Luminosity,
2$\times$2~TeV Collider in the Tevatron Tunnel,'' Report to the DPF Future
Hadron Facilities Workshop, Bloomington, Indiana, July 6-10, 1994.

\bibitem{ASST}J. Kuzminski, {\em et al}.,  ``Quench Performance of 50-mm
Aperture,
15-m-Long SSC Dipole Magnets Built at Fermilab,'' {\em Proceedings of the
XV-th International Conference on High Energy Accelerators} (1992) 588.

\bibitem{PostASST}J. Kuzminski, {\em et al}.,  ``Test of Fermilab Built, Post
ASST, 50-mm
Aperture, Full Length SSC Dipole Magnets,'' Presented at the 5th International
Industrial Symposium on the Super Collider, San Francisco, CA, 6-8 May, 1993;
SSCL-Preprint-314.

\bibitem{Strait}J. Strait, {\em et al}.,  ``Magnetic Field Measurements of Full
Length
50 mm Aperture SSC Dipole Magnets at Fermilab,'' {\em Proceedings of the
XV-th International Conference on High Energy Accelerators} (1992) 656.

\bibitem{GarrenSyphers}A.A. Garren and M.J. Syphers, ``1.8 TeV Tevatron Upgrade
Lattices,''  {\em Proceedings of the 1989 IEEE Accelerator Conference} (1989)
1364.

\bibitem{Theilacker}J. Theilacker, private communication.

\bibitem{Murphy}C.T. Murphy, private communication.

\end{thebibliography}

\end{document}